\documentclass[conference,letterpaper,onecolumn, draftclsnofoot]{IEEEtran}

\usepackage[utf8]{inputenc} 
\usepackage[T1]{fontenc}
\usepackage{url}
\usepackage{microtype}
\usepackage{ifthen}
\usepackage{verbatim}
\usepackage{cite}
\usepackage{xcolor}
\usepackage[cmex10]{amsmath}
\usepackage{amsfonts}
\usepackage{mathtools}

\newtheorem{theorem}{Theorem}[section]

\newtheorem{lemma}{Lemma}[section]

\newtheorem{example}{Example}[section]

\DeclareMathOperator{\spn}{span}
\DeclareMathOperator{\ev}{ev}

\DeclareMathOperator{\GRS}{GRS}
\DeclareMathOperator{\Rep}{Rep}

\newcommand{\F}{\mathbb{F}}
\newcommand{\enc}{\mathrm{enc}}
\renewcommand{\sec}{\mathrm{sec}}
\newcommand{\info}{\mathrm{info}}
\newcommand{\priv}{\mathrm{priv}}
\newcommand{\noise}{\mathrm{noise}}
\newcommand{\query}{\mathrm{query}}
\newcommand{\full}{\mathrm{full}}

\definecolor{pinegreen}{rgb}{0.0, 0.47, 0.44}

\begin{document}
\title{Algebraic Geometry Codes for Cross-Subspace Alignment in Private Information Retrieval}
\author{%
   \IEEEauthorblockN{Okko~Makkonen\IEEEauthorrefmark{1},
                     David~A.~Karpuk\IEEEauthorrefmark{1}\IEEEauthorrefmark{2},
                     and Camilla~Hollanti\IEEEauthorrefmark{1}}
   \IEEEauthorblockA{\IEEEauthorrefmark{1}%
                    Department of Mathematics and Systems Analysis, Aalto University, Finland,
                first.last@aalto.fi}
   \IEEEauthorblockA{\IEEEauthorrefmark{2}%
                     WithSecure Corporation,
                     Helsinki, Finland, 
            davekarpuk@gmail.com}
   
 }

\maketitle

\begin{abstract}
A new framework for interference alignment in secure and private information retrieval (PIR) from colluding servers is proposed, generalizing the original cross-subspace alignment (CSA) codes proposed by Jia, Sun, and Jafar. The general scheme is built on algebraic geometry codes and explicit constructions with replicated storage are given over curves of genus zero and one. It is shown that the proposed scheme offers interesting tradeoffs between the field size, file size, number of colluding servers, and the total  number of servers. When the field size is fixed, this translates in some cases to higher retrieval rates than those of the original scheme. In addition, the new schemes exist also in cases where the original ones do not.
\end{abstract}

\section{Introduction}

Private information retrieval (PIR) \cite{chor1995private} studies the problem of retrieving a file from a database without disclosing any information on the identity of the retrieved item. The basic variant of the problem considers public files, while security can be added to protect the contents of the files in addition to protecting user privacy. Scheme constructions for various different scenarios with related capacity results can be found in the literature, \emph{e.g.}, \cite{sun2017repcap,Sun2016,Banawan2018,tajeddine2018private,freij2017private,Tajeddine2018,freij2018tit,Oliveira2018isit,Jia_Sun_Jafar_XSTPIR,holzbaur2022tit}.

Cross-subspace alignment (CSA) codes have been recently proposed as a means to construct secure and private information retrieval schemes and secure distributed matrix multiplication (SDMM) schemes over classical and quantum channels \cite{Jia_Sun_Jafar_XSTPIR,Jia_Jafar_MDSXSTPIR, Chen_Jia_Wang_Jafar_NGCSA,Jia_Jafar_SDMM,nsumboxarxiv,QCSA23}. As the term suggests, these codes are designed such that they are capable of independently operating on different file fragments in a way that allows for the unwanted fragments (interference) to align in the overall response and cancel out, while keeping the desired information intact. This is achieved by a suitable direct sum decomposition of the ambient space.

In this work, we reinterpret the original CSA codes from \cite{Jia_Sun_Jafar_XSTPIR} as evaluation codes over the projective line.  In addition to a more conceptual construction, the resulting  algebraic-geometric interpretation admits generalization to higher-genus curves.  We then focus on genus one curves, also known as elliptic curves, and showcase the potential of genus one curves in attaining PIR rates higher than the original CSA codes for a fixed field size by allowing a small increase in the number of servers and larger files. This improvement stems from the fact that increasing the genus yields curves with more rational points. In particular, elliptic curves ($g=1$) have more rational points with respect to the field size compared to the projective line ($g=0$) making the parameter choices more flexible. This allows for interesting tradeoffs between the field size, file size, number of colluding servers, and the number of servers in total.

\section{General Framework for $X$-Secure and $T$-Private Information Retrieval}\label{sec:general_framework}

In this section, we will describe the scheme setup. For the formal definitions of the PIR rate (denoted by $\mathcal{R}$), security, and privacy of PIR protocols we refer to \cite{Jia_Sun_Jafar_XSTPIR}.

Let $A$ be an algebra over $\F_q$ and $V_\ell^\enc, V_\ell^\sec, V_\ell^\query, V_\ell^\priv \subseteq A$ be finite-dimensional subspaces for $\ell \in [L]$. In the following we define the storage system, queries and decoding of a PIR scheme by working over the algebra. We think of each of the files as consisting of $L$ fragments, \emph{i.e.}, symbols in $\F_q$. For subspaces $V$ and $W$ of $A$ we define $V \cdot W = \spn\{f \cdot g \mid f \in V, g \in W\}$.  For a positive integer $n$ we write $[n] = \{1, \dots, n\}$. 

\textbf{Storage:} For each file (indexed by $m \in [M]$) and fragment (indexed by $\ell \in [L]$) choose $f_{\ell, m}^\enc \in V_\ell^\enc$ to encode the file fragment and choose $f_{\ell, m}^\sec \in V_\ell^\sec$ uniformly at random. The servers hold $s_{\ell, m} = f_{\ell, m}^\enc + f_{\ell, m}^\sec$ for all $\ell, m$. Each file is protected with random noise from the $f_{\ell, m}^\sec$.

\textbf{Queries:} Let $\theta \in [M]$ be the index of the desired file. Let $V_\ell^\query = \spn\{ h_\ell \}$. For each file and fragment choose
\begin{equation*}
    g_{\ell, m}^\query = \begin{cases}
        h_\ell & \text{if $m = \theta$} \\
        0 & \text{otherwise}
    \end{cases}
\end{equation*}
and choose $g_{\ell, m}^\priv \in V_\ell^\priv$ uniformly at random. We query the servers with $q_{\ell, m} = g_{\ell, m}^\query + g_{\ell, m}^\priv$ for all $\ell, m$. Each query is similarly protected with noise from the $g_{\ell, m}^\priv$.

\textbf{Responses:} We receive $r = \sum_{\ell, m} s_{\ell, m} q_{\ell, m}$.  We now define the following (finite-dimensional) subspaces of $A$:
\begin{align}\label{eq:v_spaces_defn}
    V_\ell^\info &= V_\ell^\enc \cdot V_\ell^\query, \qquad
    V^\info = \bigoplus_\ell V_\ell^\info \\
    V^\noise &= \sum_\ell (V_\ell^\enc \cdot V_\ell^\priv + V_\ell^\sec \cdot V_\ell^\query + V_\ell^\sec \cdot V_\ell^\priv). \nonumber
\end{align}
where the direct sum decomposition of $V^\info$ is an assumption we place on the system.  The response $r$ can now be written
\begin{equation*}
    r = \sum_\ell \underbrace{f_{\ell, \theta}^\enc h_\ell}_{\in V_\ell^\info} + \sum_{\ell, m} \underbrace{(f_{\ell, m}^\enc g_{\ell, m}^\priv + f_{\ell, m}^\sec g_{\ell, m}^\info + f_{\ell, m}^\sec g_{\ell, m}^\priv)}_{\in V^\noise}.
\end{equation*}

\textbf{Decoding:} 
The noise terms live in the $V^\noise$ space, while the information we wish to recover lives in $V^\info$. Thus, if $V^\info \cap V^\noise = 0$, then we can recover $f_{\ell, \theta}^\enc h_\ell$ for all $\ell$ from the response $r$. This allows us to recover the file fragments for the desired file $\theta$, assuming that $h_\ell$ is a unit in the algebra.

Instead of working over the algebra $A$, the computations performed by the servers happen in $\F_q^N$ via an $\F_q$-algebra homomorphism $\varphi \colon A \to \F_q^N$, where $\F_q^N$ is equipped with the coordinatewise product (the star product). In particular, worker $n \in [N]$ stores the $n$th coordinate of $\varphi(s_{\ell, m})$ and we send them the $n$th coordinate of $\varphi(q_{\ell, m})$. The $n$th worker computes
\begin{equation*}
    \varphi(r)_n = \sum_{\ell, m} \varphi(s_{\ell, m})_n \varphi(q_{\ell, m})_n.
\end{equation*}
As long as $\varphi$ is injective on $V^\info \oplus V^\noise$ then we may still recover $r \in V^\info \oplus V^\noise$ from $\varphi(r)$.

\textbf{Privacy and security:} The noise added to the queries is chosen uniformly at random from the code $\varphi(V_\ell^\priv)$. Let $\mathcal{I} \subseteq [N]$ be a set of colluding servers. We need to show that the queries observed by $\mathcal{I}$ are independent of the desired file index~$\theta$.  If $\mathcal{C}$ is any linear code we let $d^\perp(\mathcal{C})$ denote the minimum distance of the dual code $\mathcal{C}^\perp$.

\begin{lemma}
\label{lem:privacy_and_security}
Let $G_\ell^\priv$ (resp. $G_\ell^\sec$) be a generator matrix of the code $\varphi(V_\ell^\priv) \subseteq \F_q^N$ (resp. $\varphi(V_\ell^\sec)$) and let $\mathcal{I} \subseteq [N]$ be a set such that the columns of $G_\ell^\priv$ (resp. $G_\ell^\sec$) indexed by $\mathcal{I}$ are linearly independent. The PIR scheme described above is private (resp. secure) against the servers indexed by $\mathcal{I}$ colluding. In particular, the scheme is private (resp. secure) against any $d^\perp(\varphi(V_\ell^\priv)) - 1$ (resp. $d^\perp(\varphi(V_\ell^\sec)) - 1$) servers colluding.
\end{lemma}
\begin{IEEEproof}
    The argument follows a standard proof technique in the literature; see \cite[Theorem 8]{freij2017private}.
\end{IEEEproof}

We summarize the above discussion as the following theorem.  

\begin{theorem}
\label{thm:general_framework}
Let $V_\ell^\enc, V_\ell^\sec, V_\ell^\query, V_\ell^\priv$ be subspaces of $A$ and $\varphi \colon A \to \F_q^N$ an $\F_q$-algebra homomorphism such that:
\begin{enumerate}
    \item $V_\ell^\query = \spn\{h_\ell\}$, where $h_\ell$ is a unit in $A$,
    \item $\dim(\sum_\ell V_\ell^\info) = \sum_\ell \dim(V_\ell^\info)$,
    \item $V^\info \cap V^\noise = 0$, and
    \item $\varphi$ is injective on $V^\info \oplus V^\noise$.
\end{enumerate}
Then there exists a secure and private information retrieval scheme with rate $\mathcal{R} = \frac{L}{N}$ that is $d^\perp(\varphi(V_\ell^\sec)) - 1$ secure in the $\ell$th fragment and $d^\perp(\varphi(V_\ell^\priv)) - 1$ private in the $\ell$th fragment.
\end{theorem}

The Cross-Subspace Alignment codes of \cite{Jia_Sun_Jafar_XSTPIR} allow one to set $L = N - (X+T)$, which they show results in the asymptotic capacity in this setting.  While the above result allows for varying levels of security and privacy depending on the index $\ell$ of the file fragment, we will construct our schemes such that these levels are independent of the index $\ell$.

\section{Preliminaries from Algebraic Geometry}

To use Theorem~\ref{thm:general_framework} effectively, we need to construct explicit vector spaces satisfying assumptions 1)--4) therein.  These spaces will be Riemann--Roch spaces of divisors over algebraic curves.  We review most of the necessary concepts from Algebraic Geometry below and refer to \cite[Chapters~2--4]{hoholdt} as a catch-all reference for this topic.

Throughout this section, we let $\mathcal{X}$ be be a smooth, projective curve of genus $g$ over a finite field $\F_q$ with function field $\F_q(\mathcal{X})$. 

\subsection{Divisors and the Riemann--Roch Theorem}

A \emph{divisor} $D$ on $\mathcal{X}$ is a formal sum of points
\[
D = \sum_{P\in\mathcal{X}} n_P P,\quad n_P\in \mathbb{Z}
\]
where all but finitely many $n_P$ are zero. The \emph{degree} of a divisor $D$ as above is $\deg(D) = \sum_P n_P \deg(P)$. If $D' = \sum_P n_P' P$ is another divisor on $\mathcal{X}$ we write $D\leq D'$ if $n_P\leq n_P'$ for all $P$. The set of divisors on $\mathcal{X}$ form an abelian group.

For a non-zero rational function $f \in \F_q(\mathcal{X})^*$ and a point $P \in \mathcal{X}$ we define $v_P(f) \in \mathbb{Z}$ to be the order of vanishing of $f$ at $P$. Any non-zero $f \in \F_q(\mathcal{X})^*$ defines a divisor
\[
(f) = \sum_{P\in\mathcal{X}} v_P(f) P = (f)_0 - (f)_\infty,
\]
where
\[
    (f)_0 = \!\! \sum_{P, v_P(f)>0} \!\! v_P(f)P, \quad
    (f)_\infty = \!\! \sum_{P, v_P(f)<0} \!\! -v_P(f)P
\]
are the \emph{zero divisor} and \emph{pole divisor} of $f$, respectively. 

For any divisor $D$ on $\mathcal{X}$, we can define the associated \emph{Riemann--Roch space}
\[
\mathcal{L}(D) = \{ f \in \F_q(\mathcal{X}) \mid (f) + D \geq 0 \} \cup \{0\}.
\]
This is a finite-dimensional vector space over $\F_q$, whose dimension is denoted by $\ell(D)$.

\begin{theorem}\label{thm:riemann-roch_spaces}
Let $\mathcal{X}$ be a smooth, projective curve of genus $g$ over $\F_q$, and let $D = \sum_P n_P P$ and $D' = \sum_P n_P' P$ be divisors on $\mathcal{X}$. We have the following results: 
\begin{enumerate}
    \item If $D \leq D'$, then $\mathcal{L}(D) \subseteq \mathcal{L}(D')$.
    \item If $D = -(f)$ for $f \in \F_q(\mathcal{X})^*$, then $\mathcal{L}(D) = \spn\{f\}$.
    \item If $D' = D + (h)$ then we have an isomorphism of vector spaces $\mathcal{L}(D') \rightarrow \mathcal{L}(D)$ given by $f \mapsto fh$.
    \item We have $\mathcal{L}(D) \cdot \mathcal{L}(D') \subseteq \mathcal{L}(D + D')$ with equality if $\deg(D) \geq 2g$ and $\deg(D') \geq 2g + 1$ \cite[Theorem 8]{couvreur2017cryptanalysis}.
    \item If $D$ is a divisor with $\deg(D) < 0$, then $\ell(D) = 0$.
\end{enumerate}
\end{theorem}
With the exception of 4), the above facts are all straightforward to verify from the definitions. The quantity $\ell(D)$ can be computed using the celebrated Riemann--Roch Theorem.

\begin{theorem}[Riemann--Roch Theorem]\label{thm:riemann-roch_theorem}
    There exists a divisor $K$ on $\mathcal{X}$, called the \emph{canonical divisor}, such that
    \[
    \ell(D) - \ell(K-D) = \deg(D) - g + 1
    \]
    for any other divisor $D$ on $\mathcal{X}$.
\end{theorem}

We will be most interested in the cases of $g = 0$ (the projective line) and $g = 1$ (elliptic curves). On the projective line the canonical divisor is given by $K = -2P_\infty$, and for an elliptic curve we have $K = 0$. These facts suffice to compute $\ell(D)$ for any $D$ we use in the sequel.

\subsection{Algebraic Geometry Codes}

Given a Riemann--Roch space of a divisor $D$, we may define a linear code as the image of $\mathcal{L}(D)$ under some evaluation map. In particular, let $\mathcal{P} = \{P_1, \dots, P_n\}$ be a set of $n$ rational points that are not in the support of $D$. Then
\begin{equation*}
    \ev_\mathcal{P} \colon \mathcal{L}(D) \to \F_q^n, \quad f \mapsto (f(P_1), \dots, f(P_n))
\end{equation*}
is a well-defined linear map. The algebraic geometry (AG) code corresponding to the divisor $D$ is defined as
\begin{equation*}
    \mathcal{C}(\mathcal{P}, D) = \ev_\mathcal{P}(\mathcal{L}(D)).
\end{equation*}
If $n > \deg(D)$, then $\mathcal{C}(\mathcal{P}, D)$ is an $[n, k, d]$ linear code with $k = \ell(D)$ and $d \geq n - \deg(D)$.  By the Riemann--Roch theorem, if $\deg(D) > 2g - 2$, then $\ell(D) = \deg(D) - g + 1$, so combining with the Singleton bound
\begin{equation}\label{eq:AG_singleton_bound}
    n - g + 1 \leq k + d \leq n + 1.
\end{equation}
In particular, AG codes over curves with genus $g = 0$ achieve the Singleton bound and are MDS codes. It is well-known that there exists another divisor $D^\perp$ such that
\begin{equation}\label{eq:ag_dual}
    \mathcal{C}(\mathcal{P}, D)^\perp = \mathcal{C}(\mathcal{P}, D^\perp).
\end{equation}

\begin{example}\label{ex:grs_example}
Consider the divisor $D = (k - 1)P_\infty + (h)$ on the projective line $\mathbb{P}^1$ for some non-zero rational function $h$.  Letting $P_i = [\alpha_i:1]\in\mathbb{P}^1$, setting $\nu_i = h(\alpha_i)$ and assuming $\nu_i\neq 0$, one has  $\mathcal{C}(\mathcal{P}, D) = \GRS_k(\alpha, \nu)$, a generalized Reed--Solomon code. In particular, if $D = 0$, then $\mathcal{C}(\mathcal{P}, D) = \Rep(n)$, the length $n$ repetition code.
\end{example}

\subsection{The Hasse Bound and Maximal Curves}

By a \emph{rational point} of our curve $\mathcal{X}$ defined over a field $\F_q$, we mean a solution to the equation(s) defining $\mathcal{X}$ with coordinates in $\F_q$.  The set of all rational points of $\mathcal{X}$ is denoted $\mathcal{X}(\F_q)$.  Since $\F_q$ is finite, this is a finite set.

From a coding-theoretic perspective, the advantage of passing to curves of genus $g > 0$ lies in their ability to have more than $q + 1$ rational points.  In particular, for elliptic curves over $\F_q$, that is, curves of genus $g = 1$, we have by Hasse's Theorem \cite[Theorem 4.2]{washington} that $\#\mathcal{X}(\F_q) \leq q + 1 + \left\lfloor2\sqrt{q}\right\rfloor$, and curves attaining this bound are called \emph{maximal}. We have the following existence results, cf. \cite[Theorem 4.3]{washington}.
\begin{theorem}\label{thm:elliptic_curve_points}
    Let $q = p^s$. Then there exists an elliptic curve $\mathcal{X}$ over $\F_q$ with $\#\mathcal{X}(\F_q) = q + 1 - a$ if and only if $|a| \leq 2\sqrt{q}$ and either $\gcd(a, p) = 1$, or $q$ is a square and $a = \pm 2\sqrt{q}$.
\end{theorem}

It follows easily from the above that maximal curves exist whenever $q$ is prime or a square of a prime power.
 
\section{Construction for Genus Zero}

In this section, we reinterpret the CSA codes of \cite{Jia_Sun_Jafar_XSTPIR} as algebraic geometry codes, using the language of the previous two sections.  In addition to streamlining the scheme construction and allowing for generalization to higher genus curves, this also provides simplified proofs of several of the main results of \cite{Jia_Sun_Jafar_XSTPIR}, namely the security, privacy, and decodability of the scheme; see \cite[Section 6]{Jia_Sun_Jafar_XSTPIR}.  

\subsection{Basic Scheme Outline}

To begin, we broadly outline how to construct the main subspaces of Section~\ref{sec:general_framework} so that the conditions of Theorem~\ref{thm:general_framework} are satisfied, by using algebraic geometry codes.

We let $\mathcal{X}$ be a smooth, projective curve over a finite field $\F_q$, with function field $\F_q(\mathcal{X})$. We will set $A$ to be a certain $\F_q$-subalgebra of $\F_q(\mathcal{X})$, and choose our subspaces $V_\ell^\enc$, etc.\ to be Riemann--Roch spaces of divisors on $\mathcal{X}$. In particular, $V_\ell^\enc = \mathcal{L}(D_\ell^\enc)$ for some divisor $D_\ell^\enc$, and so on.

To use Theorem~\ref{thm:general_framework} in this context, we need to construct divisors $D^\info$ and $D^\noise$ on $\mathcal{X}$ such that $\mathcal{L}(D^\info)\cap \mathcal{L}(D^\noise) = 0$, as well as find an injective map
\[
\varphi: \mathcal{L}(D^\info) \oplus \mathcal{L}(D^\noise)\rightarrow \F_q^N.
\]
The trivial intersection condition can be guaranteed by forcing functions in these spaces to have zeros and poles at prescribed locations. To construct such an injective $\varphi$, we will find a third divisor $D^\full$ such that $D^\info, D^\noise \leq D^\full$, which then forces $\mathcal{L}(D^\info)\oplus\mathcal{L}(D^\noise)\subseteq \mathcal{L}(D^\full)$. If $\mathcal{P} = \{R_1,\ldots,R_N\}$ is a set of rational points of $\mathcal{X}$ disjoint from the support of $D^\full$, then the map $\varphi = \ev_\mathcal{P}:\mathcal{L}(D^\full)\rightarrow \F_q^N$ is injective whenever $N > \deg(D^\full)$. In particular, this map will be injective when restricted to the subspace $\mathcal{L}(D^\info)\oplus\mathcal{L}(D^\noise)$.

\subsection{Explicit Scheme Construction}

To define the encoding and the security we will choose
\begin{align*}
    D_\ell^\enc &= 0 & D_\ell^\sec = (X - 1)P_\infty + (h_\ell) 
\end{align*}
for some $h_\ell \in \F_q(\mathcal{X})^*$. The AG codes corresponding to these divisors are the repetition code and, by Example~\ref{ex:grs_example}, an $X$-dimensional GRS code, respectively. For the query and privacy we choose the divisors
\begin{align*}
    D_\ell^\query &= -(h_\ell) & D_\ell^\priv = (T - 1)P_\infty
\end{align*}
such that $V_\ell^\query = \mathcal{L}(D_\ell^\query) = \spn\{h_\ell\}$, according to Theorem~\ref{thm:riemann-roch_spaces}, and the privacy code is a $T$-dimensional RS code.

According to Theorem~\ref{thm:riemann-roch_spaces}, the product of Riemann--Roch spaces corresponds to the Riemann--Roch space of the sum of the divisors. Therefore, the information space $V_\ell^\info$ for the $\ell$th fragment is associated with the divisor
\begin{equation*}
    D_\ell^\enc + D_\ell^\query = -(h_\ell).
\end{equation*}
The noise space is associated with the divisors
\begin{align*}
    D_\ell^\enc + D_\ell^\priv &= (T - 1)P_\infty \\
    D_\ell^\sec + D_\ell^\query &= (X - 1)P_\infty \\
    D_\ell^\sec + D_\ell^\priv &= (X + T - 2)P_\infty + (h_\ell).
\end{align*}
If we can find an upper bound $D^\noise$ on all of these divisors, then $V^\noise \subseteq \mathcal{L}(D^\noise)$ according to Theorem~\ref{thm:riemann-roch_spaces}.

So that the dimension of the noise space is as small as possible and does not grow with $L$, the divisor $D^\noise$ should be of minimal degree and independent of $L$ since the dimension $\ell(D^\noise)$ is essentially determined by $\deg(D^\noise)$ according to Theorem~\ref{thm:riemann-roch_theorem}. To achieve this, we will define the basis functions $h_\ell$ to all have a zero of order one at the same point, which for convenience we can choose to be $P_\infty$. This forces $(h_\ell) = P_\infty - P_\ell$ for some rational point $P_\ell \neq P_\infty$, which determines $h_\ell$ up to a non-zero multiplicative constant. So that the $h_\ell$ are all linearly independent, the points $P_\ell$ must all be chosen to be distinct. By choosing $h_\ell = \frac{1}{x - \alpha_\ell}$, where $P_\ell = [\alpha_\ell : 1]$, we have that $(h_\ell) \leq P_\infty$. Therefore, $D^\noise = (X + T - 1)P_\infty$ works as an upper bound for the noise space and $\ell(D^\noise) = X + T$.

Consider the divisor $D^\info = \sum_\ell P_\ell - P_\infty$. We see that $h_\ell \in \mathcal{L}(D^\info)$ and $\ell(D^\info) = L$. Therefore, $\{h_1, \dots, h_L\}$ is a basis of $\mathcal{L}(D^\info)$. We set $D^\full = \sum_\ell P_\ell + (X + T - 1)P_\infty$. Then,
\begin{equation*}
    \mathcal{L}(D^\info) \oplus \mathcal{L}(D^\noise) = \mathcal{L}(D^\full).
\end{equation*}
The decomposition is direct, since the functions in $\mathcal{L}(D^\info)$ have a pole outside $P_\infty$, while the functions in $\mathcal{L}(D^\noise)$ have poles only at $P_\infty$. Furthermore, $\ell(D^\full) = L + X + T = \ell(D^\info) + \ell(D^\noise)$.

Let $N = L + X + T$ and $\mathcal{P} = \{R_1, \dots, R_N\}$ be a set of rational points distinct from $P_1, \dots, P_L, P_\infty$. Then the evaluation map defined on $\mathcal{P}$ is injective on $\mathcal{L}(D^\full)$ as $N > \deg(D^\full) = L + X + T - 1$. The rate of this scheme is
\begin{equation}\label{eq:genus_0_rate}
    \mathcal{R} = \frac{L}{N} = 1 - \frac{X + T}{N}.
\end{equation}

As we need to find $N$ rational places distinct from $P_1, \dots, P_L, P_\infty$, we need that
\begin{equation*}
    \#\mathbb{P}^1(\F_q) = q + 1 \geq N + L + 1 = 2L + X + T + 1.
\end{equation*}
As for $X$-security and $T$-privacy, these properties follow easily from the apparent fact that images of $\mathcal{L}(D_\ell^\sec)$ and $\mathcal{L}(D_\ell^\priv)$ under $\varphi$ are GRS codes of dimensions $X$ and $T$, respectively.

\subsection{Interpretation in Terms of Interpolation Polynomials}

Another way to look at the construction of the basis functions $h_\ell$ is that we start with the divisor
\begin{equation*}
    D^\info = \sum_\ell P_\ell - P_\infty = -(h) + (L - 1)P_\infty,
\end{equation*}
where $h = \frac{1}{(x - \alpha_1) \cdots (x - \alpha_L)}$. Therefore, the Riemann--Roch spaces $\mathcal{L}((L - 1)P_\infty)$ and $\mathcal{L}(D^\info)$ are isomorphic through multiplication by $h$. A natural basis for $\mathcal{L}((L - 1)P_\infty)$ is $\{1, x, \dots, x^{L-1}\}$. Instead of this basis, we choose the interpolation basis, which consists of the functions
\begin{equation}\label{eq:interpolation_functions}
    \prod_{\ell' \in [L] \setminus \ell} \! (x - \alpha_{\ell'}).
\end{equation}
These functions all have the same degree (\emph{i.e.} order of pole at $P_\infty$), which means that multiplying them by $h$ results in basis functions of $\mathcal{L}(D^\info)$ which have a unique zero of order 1 at $P_\infty$.  These are the functions $h_\ell$ defined in the previous subsections.

Observe that if we had instead chosen the monomial basis of $\mathcal{L}((L-1)P_\infty)$, the order of the zero of the corresponding basis functions $h_\ell$ at $P_\infty$ would grow with $L$.  This would force the degree of the resulting divisor $D^\noise$ to grow with $L$, ultimately essentially halving the rate of the PIR scheme.

\section{Construction for Genus One}

Let $\mathcal{X}$ be an elliptic curve defined by $y^2 = f(x)$, where $f(x)$ is a cubic. In particular, we assume that $\operatorname{char}(\F_q) \neq 2, 3$.   We let $P_\infty$ denote the `point at infinity', which has projective coordinates $[0:1:0]$.  By an \emph{affine point} of $\mathcal{X}$ we mean any point $P\neq P_\infty$, which is just a pair $P = (x,y)$ satisfying the equation $y^2 = f(x)$.

\subsection{Interpolation Basis}

To proceed with the construction of our PIR scheme based on elliptic curves, we will design analogous functions $h_\ell$ based on interpolation functions as in the genus zero case.  

Let $L$ be odd and set $J = \frac{L + 1}{2}$. Let $P_1, \dots, P_J$ and $\bar{P}_1, \dots, \bar{P}_J$ be distinct affine points such that $P_j$ and $\bar{P}_j$ correspond to $(\alpha_j,\pm \beta_j)$ for some $\beta_j \neq 0$, respectively.  

We start with the divisor
\begin{equation*}
    D^\info = \sum_j (P_j + \bar{P}_j) - P_\infty = -(h) + LP_\infty,
\end{equation*}
where $h = \frac{1}{(x - \alpha_1) \cdots (x - \alpha_J)}$.  As in the genus zero case, we wish to find a basis of $\mathcal{L}(D^\info)$ consisting of function $h_\ell$ having zeros and poles in prescribed locations.  Multiplication by $h$ yields an isomorphism between $\mathcal{L}(LP_\infty)$ and $\mathcal{L}(D^\info)$, allowing us to transfer a basis from one space to the other.  

The most obvious choice for basis of $\mathcal{L}(LP_\infty)$ is
\begin{align*}
    &\{1, x, \dots, x^{J - 1}, y, yx, \dots, yx^{J-2}\} \\
    &= \{1, x, \dots, x^{J-1}\} \cup y \cdot \{1, x, \dots, x^{J-2}\},
\end{align*}
which is analogous to the basis of monomials in the genus zero case.  Instead of choosing this basis, we choose a basis whose zeros are roughly at $P_1, \bar{P}_1, \dots, P_J, \bar{P}_J$ and whose poles have roughly the same order at $P_\infty$.  Upon multiplying our functions by $h$, this will result in a basis of $\mathcal{L}(D^\info)$ whose divisors we can upper bound by a divisor whose degree is relatively small and independent of $L$.

In particular, let us choose a basis consisting of interpolation polynomials on the $\alpha_1, \dots, \alpha_J$ exactly as in \eqref{eq:interpolation_functions}, as well as these interpolation polynomials on $\alpha_1, \dots, \alpha_{J-1}$ multiplied with $y$. By multiplying these basis functions with $h$, we obtain a basis of $\mathcal{L}(D^\info)$ consisting of $\{h_1, \dots, h_L\} = \{h^{(1)}_1, \dots, h^{(1)}_J, h^{(2)}_1, \dots, h^{(2)}_{J-1}\}$, where
\begin{equation*}
    h^{(1)}_j = h \!\! \prod_{j' \in [J] \setminus j} \! (x - \alpha_{j'}), \quad
    h^{(2)}_j = h y \!\! \prod_{j' \in [J - 1] \setminus j} \! (x - \alpha_{j'}).
\end{equation*}
For example, if $L = 5$ then $J = 3$ then our basis functions are
\begin{align*}
h_j^{(1)} &= \frac{1}{x-\alpha_j} \quad \text{for $j = 1,2,3$}, \\
h_j^{(2)} &= \frac{y}{(x-\alpha_j)(x - \alpha_3)} \quad \text{for $j = 1,2$}.
\end{align*}

Our basis functions have divisors
\begin{align*}
    (h^{(1)}_j) &= 2P_\infty - (P_j + \bar{P}_j) \\
    (h^{(2)}_j) &= P_\infty + (y)_0 - (P_j + \bar{P}_j + P_J + \bar{P}_J).
\end{align*}
We can upper bound all of these divisors by $(h_\ell) \leq 2P_\infty + (y)_0$.

\subsection{Scheme Construction}

For the encoding and the security, we choose the divisors
\begin{align*}
    D_\ell^\enc &= 0 & D_\ell^\sec = (X + 1)P_\infty + (h_\ell). 
\end{align*}
For the query and the privacy we choose
\begin{align*}
    D_\ell^\query &= -(h_\ell) & D_\ell^\priv = (T + 1)P_\infty.
\end{align*}
We have that $\ell(D_\ell^\sec) = X + 1$, $\ell(D_\ell^\priv) = T + 1$, $\mathcal{L}(D_\ell^\query) = \spn\{h_\ell\}$. The information space for the $\ell$th fragment is associated with the divisor
\begin{equation*}
    D_\ell^\enc + D_\ell^\query = -(h_\ell).
\end{equation*}
The noise space is associated with the divisors
\begin{align*}
    D_\ell^\enc + D_\ell^\priv &= (T + 1)P_\infty \leq D^\noise \\
    D_\ell^\sec + D_\ell^\query &= (X + 1)P_\infty \leq D^\noise \\
    D_\ell^\sec + D_\ell^\priv &= (X + T + 2)P_\infty + (h_\ell) \leq D^\noise
\end{align*}
where $D^\noise = (X + T + 4)P_\infty + (y)_0$. We set
\begin{equation*}
    D^\full = \sum_j (P_j + \bar{P}_j) + (X + T + 4)P_\infty + (y)_0.
\end{equation*}
We have that
\begin{equation*}
    \mathcal{L}(D^\info) \oplus \mathcal{L}(D^\noise) \subseteq \mathcal{L}(D^\full).
\end{equation*}
Again, the sum is direct, since the functions in $\mathcal{L}(D^\info)$ have a pole outside $P_\infty$, while the functions in $\mathcal{L}(D^\noise)$ can only have poles at $P_\infty$.

Consider the evaluation map defined on $L + X + T + 9$ rational points not contained in the support of $D^\full$. This map is injective, since $L + X + T + 9 > \deg(D^\full) = L + X + T + 8$. The dimension of the corresponding AG code is $\ell(D^\full) = L + X + T + 8$, so there is a set $\mathcal{P}$ of $N = L + X + T + 8$ points such that the evaluation map on these points is injective. The rate of this scheme is
\begin{equation}\label{eq:genus_1_rate}
    \mathcal{R} = \frac{L}{N} = 1 - \frac{X + T + 8}{N}.
\end{equation}
As we need to find $L + X + T + 9$ rational places distinct from $P_1, \bar{P}_1, \dots, P_J, \bar{P}_J, P_\infty$ and the rational zeros of $y$, we need that the number of rational places is at least
\begin{equation*}
    \#\mathcal{X}(\F_q) \geq N + 1 + 2J + 1 + Z = 2L + X + T + 11 + Z.
\end{equation*}
where $Z \leq 3$ is the number of rational zeros of $y$.

Let $\varphi$ be the evaluation map defined on $\mathcal{P}$. Then, $\varphi(V_\ell^\priv) = \mathcal{C}(\mathcal{P}, D_\ell^\priv)$. As mentioned above, $\ell(D_\ell^\priv) = T + 1$, so this code is a $[N, T + 1]$ AG code over a genus $g = 1$ curve. The dual of this code is an $[N, N - (T + 1)]$ AG code according to \eqref{eq:ag_dual}. Therefore, according to \eqref{eq:AG_singleton_bound}
\begin{equation*}
    d^\perp(\varphi(V_\ell^\priv)) - 1 \geq N - (N - (T + 1)) - 1 = T.
\end{equation*}
Hence, the scheme is $T$-private according to Lemma~\ref{lem:privacy_and_security}. Similarly, the scheme is $X$-secure, since $\ell(D_\ell^\sec) = X + 1$.

\section{Comparison}

\begin{figure}
    \centering
    \includegraphics[width=0.6\columnwidth]{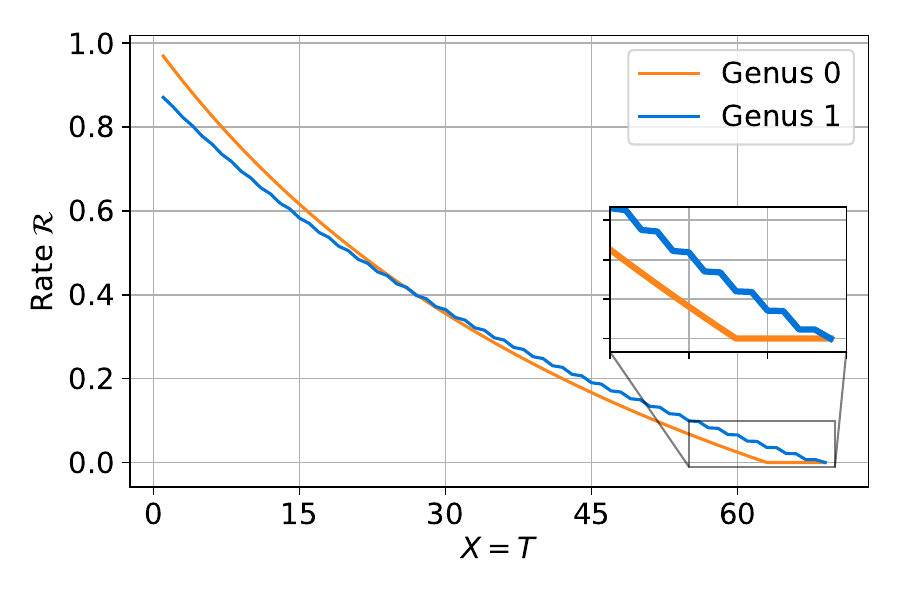}
    \caption{Comparison for $q = 127$. The genus zero curve corresponds to the original CSA scheme \cite{Jia_Sun_Jafar_XSTPIR}. The genus one construction is using the curve $\mathcal{X} \colon y^2 = x^3 + x + 33$, which has $150$ rational points and $y$ has $Z = 1$ rational zeros. The genus one scheme has a higher rate when $X = T \geq 26$.}
    \label{fig:comparison_q_127}
\end{figure}

\subsection{An Example}

Let $q = 43$ and $X = T = 16$. For the genus zero construction we need $q + 1 \geq 2L + X + T + 1$. We choose $L = 5$ and $N = 37$ as this gives the largest possible rate of $\mathcal{R} = \frac{5}{37} \approx 0.1351$. For the genus one construction we choose the curve $\mathcal{X} \colon y^2 = x^3 + 9$, which has $\#\mathcal{X}(\F_q) = 57$ rational points. Furthermore, $y$ has $Z = 0$ rational zeros. Therefore, we need $\#\mathcal{X}(\F_q) \geq 2L + X + T + 11$. We choose $L = 7$ and $N = 47$ which give a rate of $\mathcal{R} = \frac{7}{47} \approx 0.1489$, which is a $10\%$ increase in rate compared to the genus zero case.

\subsection{Discussion}

According to the rate expressions in \eqref{eq:genus_0_rate} and \eqref{eq:genus_1_rate}, the rate of the genus one scheme is always smaller than the genus zero rate for a fixed $X, T, N$, whenever both constructions exist. However, the two constructions over genus zero and genus one have different field size requirements due to the fact that curves with genus one allow for more rational points. Hence, we may utilize a larger number of servers to get a larger rate for the genus one scheme for some fixed field size. For both schemes we want to find the largest possible rate given some fixed field size $q$ and privacy/security parameters $T$ and $X$. A plot of the maximal rates for $q = 127$ is given in Figure~\ref{fig:comparison_q_127}. We see that for small values of $X = T$, the construction with genus zero achieves a better rate, but for $X = T \geq 26$, the construction with genus one achieves a better rate. Furthermore, for even larger values of $X = T$, the construction with genus zero does not exist due to the field size constraint, while the construction with genus one can still be achieved.

Both of the constructions were designed to be $T$-private and $X$-secure. This means that the schemes are private (resp. secure) against any $T$ (resp. $X$) colluding servers. However, for the genus one construction the privacy and security codes $\varphi(V_\ell^\priv)$ and $\varphi(V_\ell^\sec)$ are $T + 1$ and $X + 1$ dimensional, respectively. Therefore, there are some sets of servers of size $T + 1$ (resp. $X + 1$) such that the scheme is private (resp. secure) against these colluding servers.

\section{Conclusions and Future Work}

We have proposed a new framework for interference alignment in secure and private information retrieval from replicated and colluding servers.  Explicit constructions from algebraic geometry codes were given over curves of genus zero and one. We demonstrated improved PIR rates using codes over elliptic curves compared to the original CSA scheme, when the field size is fixed and the number of servers and the file size are allowed to vary. Future work consists of describing the scheme for coded storage as well as for higher genus curves.

\section*{Acknowledgment}
This work has been supported by the Research Council of Finland under Grant No.\ 336005 and by the Vilho, Yrjö and Kalle Väisälä Foundation of the Finnish Academy of Science and Letters.

\bibliographystyle{IEEEtran}
\bibliography{bib}

\end{document}